\documentclass[a4paper,11pt]{article}
\usepackage{pos}

\title{The accretion-driven eruption of the recurrent nova \\ T Corona Borealis}
\ShortTitle{The accretion-induced eruption of T~CrB}

\author*[a]{Raymundo Baptista}
\author[b,c]{Wagner Schlindwein}
\author[d,e]{Gerardo J. M. Luna}

\affiliation[a]{Departamento de F\'{i}sica,
  Universidade Federal de Santa Catarina, \\
  Campus Trindade, Florian\'{o}polis, SC, Brazil}

\affiliation[b]{Instituto Nacional de Pesquisas Espaciais - INPE,\\
  Avenida dos Astronautas, 1758, São José dos Campos-SP, Brazil}

\affiliation[c]{Departamento de F\'{i}sica,
Universidade Estadual de Maring\'a - UEM, \\
Av.\ Colombo, 5790, Maringá-PR, Brazil}

\affiliation[d]{Universidad Nacional de Hurlingham (UNAHUR),
    Secretaría de Investigación, \\
    Av.\ Gdor.\ Vergara 2222, Villa Tesei, Buenos Aires, Argentina}

\affiliation[e]{Consejo Nacional de Investigaciones
Científicas y Técnicas (CONICET).}

\emailAdd{raybap@gmail.com}
\emailAdd{wagner.schlindwein@astro.ufsc.br}
\emailAdd{juan.luna@unahur.edu.ar}

\abstract{T~Corona Borealis (T~CrB) is a symbiotic recurrent nova with
an $\simeq 80$\,yr recurrence interval, the eruptions of which occur on
top of a $\simeq 15$\,yr long high-brightness state. We show that the
high-brightness state is best explained as the response of a high-viscosity
($\alpha=3$) accretion disk to a unique event in which the mass transfer
rate from the donor star increases by a factor $\simeq 100$, from
$\dot{M}\mathrm{(quies)}= 2 \times 10^{-9}\, M_\odot$\,yr$^{-1}$ up to
$\dot{M}\mathrm{(out)}= 1.9 \times 10^{-7}\, M_\odot$\,yr$^{-1}$; it can
not be a thermal-viscous disk instability outburst neither a steady
nuclear burning event.
The constraint that the matter accreted onto the white dwarf in between
eruptions equals the envelope mass $M_{ig}$ needed to trigger nova
eruptions at the observed recurrence interval requires a white dwarf mass
of $M_1= 1.29\,M_\odot$, a donor star mass of $M_2= 0.7\,M_\odot$, and an
inclination of $i= 57.3^o$. 
As the high-brightness state responds for 95\% of $M_{ig}$, the nova
eruptions of T~CrB are induced by accretion events. Without the 15\,yr
long enhanced mass transfer events, its nova recurrence interval would
be significantly longer, $\simeq 5500$\,yr. 
T~CrB exhibits a conspicuous decrease in brightness during the 1-2\,yr
prior to the nova event. We argue that this pre-eruption dip occurs
during the convection phase that precedes the nova eruption and is best
explained by the slow, accelerated expansion of the accreted envelope
(and inner disk radius) at an average velocity of $v_\mathrm{exp}=
0.02$\,km\,s$^{-1}$ over a 2\,yr timescale, likely as a consequence of
excess heat being increasingly deposited at the accreted layer by 
thermonuclear reactions before the nova eruption stage.}

\FullConference{The Golden Age of Cataclysmic Variables and Related Objects - VII (GOLDEN2025)\\
1-6 September, 2025\\
Mondello, Palermo, Italy\\}

\begin{document}
\maketitle

\section{Context and motivations}
\label{context}

\begin{figure}
\center
\includegraphics[width=0.6\columnwidth]{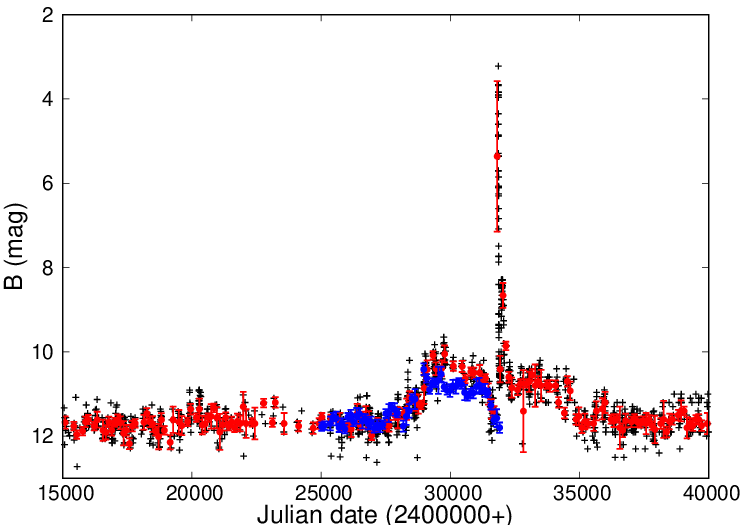}
\caption{$B$-band light curve of T~CrB during the 1938-1955
high-brightness state and the associated 1946 nova eruption (black
crosses). Red dots with error bars show the same data after median
filtering with a running box of width 113\,d. The recent AAVSO data
from 2004 until April 2024 are shown in blue, after being processed
with the same median filter and shifted in time by -78\,yr to match
the 1946 data. From \citep{sbl25}.
\label{fig:dados}}
\end{figure}

T~CrB is the nearest symbiotic recurrent nova. It contains an M4\,III
giant (RG) \citep{Zamanovet2023} which fills its Roche lobe and
transfers mass to a massive white dwarf (WD) via an accretion disk of
radius $R_d=89 \pm 19 \,R_\odot$ \citep{Zamanovet2024}. Twice in the
last two centuries, in 1866 and 1946, the accreted material ignited
on the surface of the WD via runaway thermonuclear fusion reactions
(TNR) and produced a nova eruption, suggesting a recurrence
time of $T_R = (80\pm 2)$\,yr \citep{Schaefer2023a}.
Figure~\ref{fig:dados} shows the recent optical behavior of T~CrB,
overlaid on the light curve from around the time of the 1946 eruption
but shifted by -78 years. The alignment between both light curves is
remarkable and most likely the current high-accretion state would lead
to the next nova eruption, as did in the two previous eruptions. The
comparison between current and previous brightness behavior has allowed
different authors to predict that the next nova eruption will occur
within 2025$\pm$2\,yr \citep{Lunaet2020,Schaefer2019,Zamanovet2023}.

One challenging aspect of T CrB is the fact that both eruptions occurred
approximately midway through a transient state of high luminosity which
lasts for $\simeq 15$\,yr \citep{Schaefer2010,Schaefer2023a}. A possible
explanation of such a state is a dwarf-nova-like outburst, which may
arise from a transient increase in the mass transfer rate of the donor
star. There is mounting evidence that this increase in brightness is
due to an increase in the accretion rate
\citep{Lunaet2018,Schaefer2023a,Ilkiewiczet2023}: UV observations in
quiescence lead to an accretion rate of $2.0\times 10^{-9} \,
M_\odot$\,yr$^{-1}$ \citep{Selvelliet1992,Stanishevet2004},
while x-ray observations lead to an estimated lower limit to the
accretion rate during the high-brightness state of $8.5 \times 10^{-8} 
\, M_\odot$\,yr$^{-1}$ \citep{Lunaet2018}. Furthermore, the
spectroscopic analysis of the recent high-brightness state by
\citet{Planquartet2025} demonstrates that this state correlates with
the bright spot enhancement, and that the increase in bright spot
luminosity must originate from either a decrease in the outer disk
radius or an increase in the mass transfer rate. Another piece of
the T~CrB puzzle is the significant decline in brightness that occurs
1-2\,yr before the nova eruption and brings the system to a brightness
similar to its quiescence level \citep{Munari2023,Zamanovet2023} (this
phase has now been coined as the pre-eruption dip \citep{Schaefer2010}).

Given the recurrent nova nature of T~CrB, the ignition of the following
nova eruption requires the previous accumulation of a hydrogen-rich
envelope of accreted matter, $M_{ig}$, onto its WD surface
\citep[e.g.,][]{Fujimoto1982,ShenBildsten2009,Wolfetal2013}. $M_{ig}$
depends on $M_1$ and on the WD mass accretion rate \citep[see][]
{ShenBildsten2009}. A larger WD mass leads to higher accreted
envelope temperatures and densities, which allows for the ignition
conditions to be reached at lower envelope masses, whereas higher
accretion rates lead to higher compressional heating, also resulting
in ignition occurring at lower envelope masses.

In the case of T~CrB, the envelope mass required to trigger a nova
eruption onto a WD of mass $M_1=1.3\,M_\odot$ fed by a solar composition
gas at a quiescent accretion rate of $2\times 10^{-9}\,M_\odot$\,yr$^{-1}$
is $M_{ig}\simeq 1.1\times 10^{-5} \, M_\odot$ \citep{ShenBildsten2009}.
The timescale required to accumulate this envelope mass at the quiescent
accretion rate is $\simeq 5500$\,yr, almost two orders of magnitude longer
than the observed $\approx$80\,yr recurrence time. This implies that {\em
there must be a phase of significantly enhanced mass accretion} which
reduces $M_{ig}$ and shortens the timescale required to trigger the
nova eruption down to the observed recurrence time \cite{sbl25}. Indeed,
at a higher mass accretion rate of $\simeq 1.7 \times 10^{-7} \,
M_\odot$\,yr$^{-1}$ the required envelope mass is reduced to
$M_{ig}\simeq 2.6\times 10^{-6} \,M_\odot$ and the ignition conditions
can be reached after only $\simeq 15$\,yr, which is in good agreement
with the full length of the high-brightness state. This reasoning
suggests that a high-mass accretion phase is key to explain the
timings and recurrent nova behavior of T~CrB, and sets the
motivation for the present work.

Here, we report the modeling of the observed high-brightness state as
an event of enhanced mass transfer rate from the RG, with the additional
requirement that the matter accreted over a time length of $(80\pm 2)$\,yr
equals the required envelope mass needed to trigger nova eruptions at the
observed recurrence interval. The model and its results are presented in
Sect.\,\ref{model}. The results are discussed in Sect.\,\ref{discuss}
and a summary of the conclusions is presented in Sect.\,\ref{summary}.

\section{Constraints \& model}
\label{model}

The Mass Transfer Outburst (MTO) model assumes that dwarf nova
outbursts are the response of a high-viscosity disk ($\alpha \geq 1$) 
\footnote{We adopt the prescription of \citet{ss73} and write the disk
  viscosity as $\nu= \alpha\,c_s H$, where $c_s$ is the local sound
  speed and $H$ is the vertical scaleheight.}
to a transient of enhanced mass transfer from the donor star
\citep{bath,BathPringle81}.
The MTO simulation code of \citet{SB2024} was used to model the
observed high-brightness state of T~CrB with the assumption that
it is a unique, 15\,yr-long enhanced mass transfer event.

For this purpose, the historical dataset of T~CrB covering the epoch
before and after the 1946 nova eruption \citep{Schaefer2023b} were
combined with archive $BVRI$-band observations of the American
Association of Variable Star Observers (AAVSO) from early 2004 until
April 2024 (Fig.\,\ref{fig:dados}). In order to retain only the long-term
trend, we smoothed the historical and the AAVSO light curves by applying
a median filter with a width of 113\,d (half the orbital period; red and
blue points with error bars in Fig.\,\ref{fig:dados}, respectively). The
historical data include both the brightening that started around 1938
and the nova eruption that occurred in 1946, while the AAVSO data
include the current brightening event and were shifted by -78 years to
match the start of the current brightening with that of the 1938 event.

The MTO model assumes a quiescent disk radius of $R_d = 0.74\,R_{L1}$
  \footnote{$R_{L1}$ is the distance from the WD to the inner
   Lagrangian point $L1$.}, 
consistent with that inferred by \citet{Zamanovet2024}, local blackbody
emission, and an enhanced mass transfer event with shape described by,
\begin{equation}
\dot{M}_2 = \dot{M}_2^q + (\dot{M}_2^h-\dot{M}_2^q) \,
 \exp \left[ -\frac{1}{2} \left( \frac{t-t_e}{b \, \Delta t_e} \right)^n \right] \, ,
\label{eq:pulso}
\end{equation}
where $\dot{M}_2^q$, $\dot{M}_2^h$, $t_e$ and $\Delta t_e$ are the
quiescent and high-accretion state mass transfer rates, the event center
time, and its full width at half maximum (FWHM), respectively. The
(integer) parameter $n$ in the exponential argument is used to adjust
the length of the plateau and the parameter $b=[2(2 \ln 2)^{1/n}]^{-1}$
is added to ensure that $\Delta t_e$ is its FWHM. The best-fit model
has $n=8$. An M4~III spectral type was assumed for the RG
($V$ magnitude from \citet{Zamanovet2023} and color indexes from
\citet{Allerbook1996}). The RG is by far the dominant light source
in the $V$, $R$ and $I$ passbands during quiescence. The irradiation
effect of the accretion luminosity onto the RG \citep{Hameuryet2020}
and interstellar extinction effects
\citep{Cardelliet1989,SchlaflyFinkbeiner2011} were taken into account;
the contribution of the WD emission to the $BVRI$ passbands is
considered negligible. The $t_e$ and $\Delta t_e$ values are inferred
directly from the historical light curve, $\alpha$ is derived from
the observed transient decline timescale, while $\dot{M}^q_2$ and
$\dot{M}_2^h$ are obtained by fitting the $BVRI$ quiescent and
high-brightness levels, respectively \cite{sbl25}.

\begin{figure}
\centering
\includegraphics[width=0.6\columnwidth]{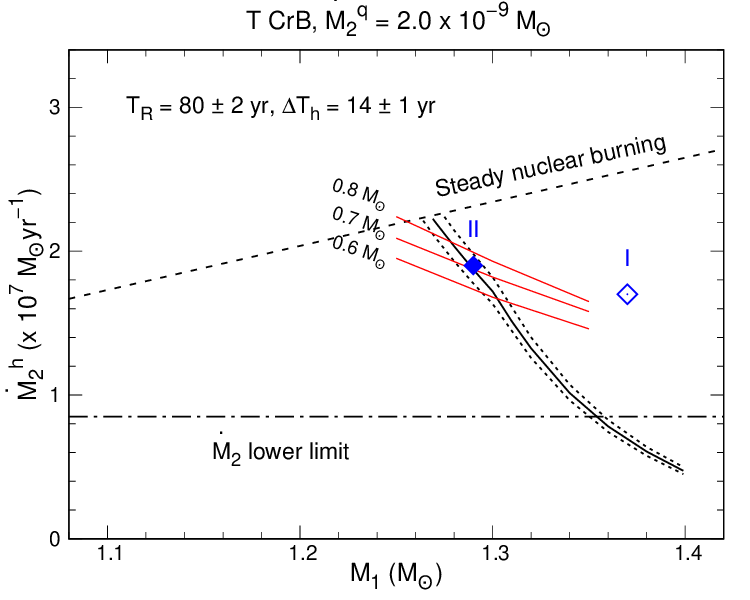}
\caption{Diagram of accretion rate versus WD mass for T~CrB. The upper
black dashed line represents the limit above which steady nuclear burning
occurs. The lower limit in $\dot{M}$ \citep{Lunaet2018} is indicated by
the black dashed-dotted line. The black solid and dotted lines mark the
relationship $\dot{M} (M_1)$ from the envelope mass that produces a nova
eruption, for an assumed nova eruption recurrence interval of $T_R = 80
\pm 2$\,yr and for a time spent in the high-accretion state of $\Delta
T_h = 14 \pm 1$\,yr. The red solid lines indicate the relationship
$\dot{M} (M_1)$ adopting $M_2 = 0.6$, 0.7 and 0.8\,$M_\odot$ in the
T~CrB primary mass function of \citet{Fekelet2000}. Open and filled
blue diamonds mark the solutions for models I and II, respectively.
From \citep{sbl25}.
\label{fig:diagrama}}
\end{figure}

The MTO model includes the additional constraint that the matter
accreted by the WD in between eruptions equals the required envelope
mass needed to trigger nova eruptions $M_{ig}$ at the observed
recurrence interval of $80\pm2$\,yr. A diagram of the high-accretion
mass transfer rate $\dot{M}_2^h$ required to reach the corresponding
$M_{ig}$ value at the end of the recurrence interval is plotted as a
function of WD mass in Fig.\,\ref{fig:diagrama}. The relationship
$\dot{M} (M_1)$ was estimated using curves from \citet{ShenBildsten2009},
assuming a recurrence interval of $80 \pm 2$\,yr, a time spent at the
high-accretion rate state of $14 \pm 1$\,yr, and a quiescence accretion
rate of $2.0 \times 10^{-9}\, M_\odot$\,yr$^{-1}$ \citep{Lunaet2018}.
The limit above which steady nuclear burning occurs, i.e., there no
longer are recurrent nova eruptions, comes from the expression of
\citet{Warner2003}; the lower limit for $\dot{M}$ comes from
\citet{Lunaet2018}. Consistent model solutions are along the solid line;
solutions below and to the left (above and to the right) of this line
imply recurrence intervals longer (shorter) than 80\,yr. With this
constraint in mind, the range of possibilities for $M_1$ becomes
reasonably narrow, between 1.27 and 1.35\,$M_\odot$. Lower $M_1$ values
put the system in the steady nuclear burning region with no recurrent
nova events, while higher $M_1$ values lead to nova recurrence intervals
shorter than observed. The reader is referred to \citet{sbl25} for
additional details about the MTO modeling.

\begin{figure}
\centering
\includegraphics[width=0.6\columnwidth]{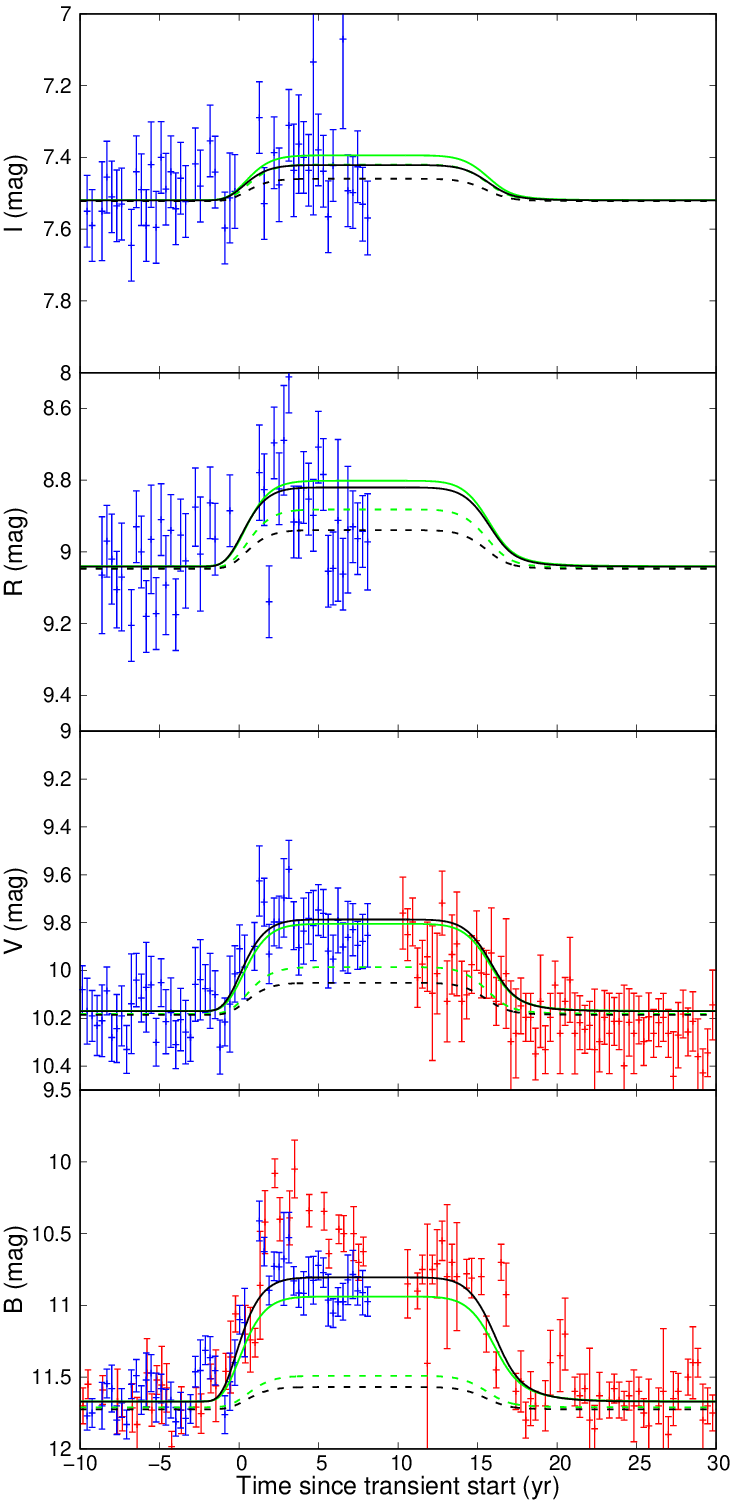}
\caption{Simulation results with model~I (green solid curve) and model~II
(black solid curve) parameters. The points with error bars are the same
as in Fig.\,\ref{fig:dados} for the $B$-band and analogous for the other
bands. The green and black dashed curves are the contribution from the
irradiated RG companion to models~I and II, respectively. From
\citep{sbl25}.
\label{fig:simulacao}}
\end{figure}

A solution using binary parameters from the literature (model\,I,
$M_1=1.37\,M_\odot$, $M_2=1.12\,M_\odot$, $i=67^o$ \citep{Stanishevet2004}
and $\dot{M}_2^h= 1.7\times 10^{-7}\, M_\odot$\,yr$^{-1}$) is not
self-consistent because it implies a recurrence interval shorter than
observed. By constraining the solution to fall along the self-consistency
solid line in Fig.\,\ref{fig:simulacao} we obtain best-fit model\,II,
with $M_1 = 1.29 \, M_\odot$, $M_2= 0.7 \, M_\odot$, $i = 57.3^\circ$,
high-accretion state mass transfer rate of $\dot{M}_2^h = 1.9 \times
10^{-7}\, M_\odot$\,yr$^{-1}$, and viscosity parameter $\alpha=3$.
Model\,II is marked as a blue filled diamond in Fig.\,\ref{fig:diagrama};
its best-fit $BVRI$ light curves are shown in Fig.\,\ref{fig:simulacao}.
It provides a better description of the observations than model\,I
while consistently falling onto the $\dot{M}(M_1)$ relationship of
Fig.\,\ref{fig:diagrama}.

\section{Discussion}
\label{discuss}

\subsection{Are there alternative explanations for the high-brightness
state?}
\label{alternative}

One possible explanation for the 15\,yr long high-brightness state of
T~CrB is to consider it the consequence of a dwarf-nova type outburst
powered by the thermal-viscous disk instability (DI) model.
DI assumes that matter accumulates in a low-viscosity disk ($\alpha_c
\sim 0.01$) during quiescence, with the surface density progressively
increasing everywhere until a critical limit is reached at some radius,
bringing the disk into a higher viscosity ($\alpha_h \sim 0.1-0.2$)
regime where the accumulated matter quickly accretes onto the WD
\citep{Lasota2001}. The mass transfer rate, $\dot{M}_2$, remains
constant during the whole outburst cycle.

Taking $\dot{M}_\mathrm{low}= 2\times 10^{-9} \, M_\odot$\,yr$^{-1}$
as the residual mass accretion onto the WD in quiescence, we estimate
$\dot{M}_2$ from the envelope mass that needs to accumulate onto the
WD surface in order to trigger the next nova eruption and from the
recurrent nova timescale, we infer the effective rate at which matter
accumulates in the disk as $\dot{M}_\mathrm{accum}= \dot{M}_2 - 
\dot{M}_\mathrm{low}= 3 \times 10^{-8} \, M_\odot$\,yr$^{-1}$, and 
we estimate the total mass accumulated in the disk over a quiescence
interval of $T_q= (65\pm 3)$\,yr as $\Delta M_\mathrm{disk}= (1.95
\pm 0.09)\times 10^{-6} M_\odot$. For an outburst mass accretion rate
of $\dot{M}_\mathrm{out}= \dot{M}_2^h= 1.9\times 10^{-7} \,
M_\odot$\,yr$^{-1}$, it takes $T_\mathrm{out}= (10.3\pm 0.5)$\,yr to
dump the accumulated matter onto the WD. This predicted DI outburst
duration is significantly shorter than the overall length of the
high-brightness state at the $5\sigma$ confidence level, indicating
that not enough mass could be accumulated in the accretion disk during
the quiescence period to sustain a DI outburst for the observed duration.
Furthermore, the nova eruption occurs roughly half-way the 
high-brightness state, probably ejecting not only the WD envelope mass
but also the whole accretion disk mass in the process. Given that the
high-brightness state resumes after the end of the T~CrB nova
eruption, the maximum amount of disk mass that could accumulate after
the disk disruption (assuming that the disk starts refilling just
after nova maximum) would account for an outburst duration (at the
$\dot{M}_\mathrm{out}$ rate) of $\leq 60$\,d, two orders of magnitude
shorter than the observed 7\,yr length of the high-brightness state
after the nova eruption. This makes the DI scenario a non-viable
explanation for the high-brightness state in T~CrB \citep{sbl25}.

The alternative explanation of the high-brightness state as the
consequence of a steady nuclear burning phase at the WD surface also has
several problems. First, in order for steady nuclear burning to occur at
the surface of a $M_1= 1.29\,M_\odot$ WD the mass accretion rate needs to
be $\dot{M}_\mathrm{burn} \geq 2.3\times 10^{-7} \, M_\odot$\,yr$^{-1}$
(see Fig.\,\ref{fig:diagrama}), $\geq 20$\% larger than inferred from
the modeling of the observations. The resulting amplitude of the 
high-brightness state as well as the magnitude of the irradiation effect
would be systematically larger than observed. Second, the predicted
luminosity from steady H-burning at the surface of a $M_1= 1.29\,M_\odot$
WD \citep{Wolfetal2013} is a factor of $\simeq 30$ larger than the
inferred luminosity of the accretion disk at the high-brightness state,
implying that T~CrB should be $\simeq 3.7$\,mag brighter than observed
in the blue. Finally, the transition between the steady and unsteady
nuclear burning regimes is very sharp, making these two possibilities
mutually excludent \citep{Paczynski1983,Wolfetal2013}: steady nuclear
burning prevents the nova eruption to occur -- whereas in T~CrB the
nova eruption is superimposed on the high-brightness state. All these
discrepancies make the steady nuclear burning scenario also a non-viable
explanation for the high-brightness state in T~CrB \citep{sbl25}.

\subsection{What is the cause of the pre-eruption dip?}

If the high-brightness state is a unique 15\,yr long event of enhanced
mass accretion around the nova eruption, what explains the 
$\simeq 1$-2\,yr long pre-eruption dip?

Recent observations indicate that the pre-eruption brightness decline
of T~CrB is much more pronounced in the blue than in the red
\citep{Munari2023,Stoyanovetal2024}, and that the half width at zero
intensity (HWZI) of the Balmer emission lines decrease from
$\mathrm{HWZI} \simeq 500$\,km\,s$^{-1}$ in the high-brightness state
\citep{Munariet2016,Zamanovet2024} down to to $\mathrm{HWZI} \simeq
(230-250)$\,km\,s$^{-1}$ around the minimum of the pre-eruption dip
(on 2023 October \citep{Zamanovet2024}). In addition, the lack of
evidence for enhanced IR emission \citep{Woodwardetal2023} and the
fact that the amplitude of the ellipsoidal modulation from the
distorted RG remains constant along the brightness decline phase
\citep{Schaeferetal2023} indicates that it is the accretion disk which
is fading in brightness by a factor of $\simeq 100$ (in order to bring
it back close to its quiescent brightness level).

There are two possible ways to decrease the accretion disk luminosity:
one is by decreasing the mass transfer rate and the other is by
increasing the inner disk radius. A reduction in mass transfer rate
by a factor $\simeq 100$ results in brightness variations of
approximately the same magnitude across the $BVRI$ passbands, while
leading to a significant increase in Balmer emission lines width to
$v_\phi \leq 1050$\,km\,s$^{-1}$ (as the systematic reduction of
accretion disk temperatures at all radii would allow the inwards
extension of the hydrogen line emitting region down to $R \geq
0.22\,R_\odot$). Both predictions are in contradiction with the
observations.

On the other hand, an increase in inner disk radius by the same factor
leads to a color-dependent brightness variation, more pronounced in
the $B$-passband and of progressively lower amplitude with increasing
wavelength -- in good agreement with the color changes observed during
the pre-eruption dip. This is the consequence of the fact that increasing
the inner disk radius selectively eliminates the hottest and bluest inner
disk regions. In addition, increasing the inner disk radius by a factor
$\simeq 100$ also increases the volume of a geometrically thin,
equatorial boundary layer by a comparable amount, allowing it to become
optically thin during the pre-eruption dip -- again in good agreement
with the observations \citep{Stoyanovetal2024}. Furthermore, the
observed HWZI at the minimum of the pre-eruption dip corresponds
to an inner disk radius for hydrogen emission of $R_{in}\simeq
(3.9-4.6)\,R_\odot$, a factor $\simeq 20$ times larger than predicted
by the decreasing mass transfer rate scenario. These results suggests
a significant increase in inner disk radius from the high-brightness
state to the pre-eruption dip \citep{sbl25}.

We further tested this idea by simulating the expansion of the inner
disk radius of model\,II over a 2\,yr time interval, keeping the mass
transfer rate $\dot{M}^h_2$ fixed, from an initial value of $R_{in}= 
R_{WD}= 0.0045\,R_\odot$ up to a few $R_\odot$, in two scenarios:
(i) at a constant acceleration, corresponding to an average expansion
velocity $v_\mathrm{exp}$, and (ii) at a constant expansion velocity.
We find that the observed brightness decline in the $BVRI$ passbands
is reasonably well reproduced by the model with accelerated inner
disk expansion, and that the best-fit to the observations is
obtained for a small, average expansion velocity of
$v_\mathrm{exp}= 0.02$\,km\,s$^{-1}$ \cite{sbl25}.

The above observational picture indicates that the decline in accretion
disk luminosity associated to the pre-eruption dip is not caused by a
decrease in mass transfer rate but by a significant, slow increase in
inner disk radius during this time. What could cause the slow inner
disk radius expansion?

The sequence of events that leads to a nova eruption can be described by
three phases \citep[e.g.,][]{Prialnik1986,ShenBildsten2009,Wolfetal2013}.
The pre-nova {\em accretion phase} ends when the accumulated envelope
mass reaches the ignition conditions. The {\em convection phase} starts
with the onset of nuclear reactions at the base of the accreted envelope.
Since the gas is degenerate, the H-burning region spreads rapidly both
inwards and outwards, the nuclear luminosity $L_\mathrm{nuc}$ rises
exponentially, temperature gradients increase and the H-burning region
becomes convectively unstable. The timescale for the development of
convection throughout the accreted envelope (i.e., the duration of the
convection phase) is short compared to the thermal timescale. Therefore,
the envelope is out of thermal equilibrium; its bolometric luminosity
remains almost constant ($L_\mathrm{bol}\sim 10^{-2}\,L_\odot$) while
$L_\mathrm{nuc}$ increasingly exceeds $L_\mathrm{bol}$ along this phase
\citep[e.g.,][]{Prialnik1986}. The end of the convection phase sets
the start of the {\em explosion phase}, with the TNR on a fully
convective envelope leading to its fast expansion and significant
increase in luminosity identified as the nova eruption. The duration of
the convection phase is 20-100 times shorter than that of the preceding
accretion phase \citep{Prialnik1986,ShenBildsten2009}. For T~CrB, the
accretion phase of which lasts about 80\,yr, this leads to an estimated
length of $(0.8-4.0)$\,yr, in good agreement with the observed $\simeq 
(1-2)$\,yr duration of the pre-eruption dip. Hence, the pre-eruption
dip occurs during the convection phase in T~CrB.

We suggest that the inferred expansion of the inner disk radius is
connected to the lack of thermal equilibrium and the development of strong
convection in the accreted envelope during this phase. We first note that,
at the onset of the convection phase on top of a high-mass WD fed at high
accretion rates, the gas is partially degenerate even at the base of the
envelope ($T_b\simeq 3\times 10^7\,K$ and $\rho_b\simeq 10^3$\,g\,cm$^3$  \citep[e.g.,][]{ShenBildsten2009}) and gas pressure is sensitive to
temperature. According to the first law of thermodynamics, the additional
heat provided to each layer across the accreted envelope drives changes
in their temperature and density at rates \citep{Kippenhahn},
\begin{equation}
  \frac{dq}{dt} = c_\rho \frac{dT}{dt} + c_T \frac{d\rho}{dt}, 
  \,\,\,\,\,\,
   c_\rho= \left(\frac{\partial u}{\partial T} \right)_\rho > 0, 
   \,\,\,\,\,\,
   c_T = \left(\frac{\partial u}{\partial\rho}\right)_T\!- \frac{P}{\rho}
   < 0 \, ,
\end{equation}
where $c_\rho$ and $c_T$ are, respectively, the specific heat at constant
density and temperature, $u$ is the internal energy per mass and $P$ is
the total pressure. A $dq/dt>0$ implies $dT/dt>0$ (because $c_\rho>0$)
and $d\rho/dt<0$ (because $c_T<0$). Thus, the increasing excess heat
deposited at the accreted envelope during the convection phase
continuously increases temperature (leading to pressure unbalance, loss
of hydrostatic equilibrium, and expansion) and decreases density (i.e.,
direct expansion), driving the envelope into slow (but accelerated)
expansion at roughly constant bolometric luminosity while convection
spreads and efficiently transports excess nuclear energy further upwards.
If the inner disk radius coincides with the surface of the accreted
envelope, the expansion of the envelope leads to the progressive
increase of the inner disk radius and to the continuous decrease of the
accretion disk luminosity even if the mass accretion rate remains
constant during the process. The reader is referred to \citet{sbl25}
for a comprehensive discussion about this idea.

\subsection{Which came first, the chicken or the egg?}

\citet{Schaefer2023b} raised the interesting issue that ``The
existence of the pre-eruption high-state presents a mystery as to why
the system brightens before the eruption", and then concluded that
``So we are left with no explanation for the enigma as to why the
high-state anticipates the eruption by 10\,yr, just as we have no
explanation for the cause or physical mechanism of the high-state
existence". This apparent enigma is a consequence of an inversion of
the relevance of the events in T~CrB \citep{sbl25}.

As remarked in Sect.\,\ref{context}, without a phase of enhanced mass
accretion, T~CrB would be a classical nova with a much longer interval
between successive eruptions ($\simeq 5500$\,yr). The best explanation
for the high-brightness state (Section~\ref{alternative}) is that it
corresponds to the crucial phase of enhanced mass accretion which
determines the much shorter recurrence interval of the T~CrB nova
eruptions. It is thus correct to say that the TNRs of T~CrB are powered
by accretion events. Indeed, the matter accumulated onto the WD surface
during the 15\,yr long enhanced accretion phase responds for 95\% of
$M_{ig}$, indicating that the probability that the ignition conditions
be reached (and lead to the nova eruption) during the high-brightness
state is also of 95\%. This provides a natural explanation for why the
nova eruption is superimposed on the high-brightness state. In other
words, it is the increased mass accretion phase during the high-brightness
state (the cause) that drives the nova eruption (the consequence)
\citep{sbl25}, underscoring previous suggestions by \citet{Lunaet2020}
and \citet{Zamanovet2023}, and solving the apparent enigma raised by
\citet{Schaefer2023b}.

\section{Conclusions}
\label{summary}

This contribution can be summarized as follows:
\begin{itemize}
    \item The T~CrB high-brightness state is best explained as a 
    mass-transfer outburst (MTO) of duration $\Delta t= 15$\,yr onto
    a high-viscosity ($\alpha= 3$) accretion disk. It can not be
    a thermal-viscous disk instability outburst neither a steady 
    nuclear burning event.

    \item The self-consistent model of the MTO event ensures that the
    matter accreted by the WD in between eruptions equals the required
    envelope mass needed to trigger nova eruptions, $M_{ig}$, 
    at the observed recurrence interval of $80\pm2$\,yr, and requires
    a WD mass of $M_1= 1.29\,M_\odot$, a RG mass of 
    $M_2= 0.7\,M_\odot$, an inclination of $i= 57.3^o$, and quiescent
    and outburst mass transfer rates of $\dot{M}\mathrm{(quies)}= 2
    \times 10^{-9}\,M_\odot$\,yr$^{-1}$ and $\dot{M}\mathrm{(out)}=
    1.9 \times 10^{-7}\,M_\odot$\,yr$^{-1}$, respectively.

    \item The high-brightness state responds for 95\% of $M_{ig}$,
    indicating that the nova eruptions of T~CrB are induced by accretion
    events. Without the 15\,yr-long MTO events, the T~CrB nova recurrence
    interval would be significantly longer, at $\simeq 5500$\,yr.

    \item The pre-eruption dip occurs during the convection phase that
    precedes the TNR and is best explained by the slow, accelerated
    expansion of the inner disk radius at an average velocity of
    $v_\mathrm{exp}= 0.02$\,km\,s$^{-1}$ over a 2\,yr timescale, likely
    as a consequence of the excess heat being increasingly deposited at
    the accreted layer by nuclear reactions before the TNR stage.
    
\end{itemize}

\acknowledgments
R.\ Baptista acknowledges financial support from CNPq/Brazil
grant 421034/2023-8. WS acknowledges financial support from
CNPq/Brazil grants 300834/2023-3, 301366/2023-3, 
300252/2024-2 and 301472/2024-6. GJML is member of the
CIC-CONICET (Argentina).


\begin{thebibliography}{99}

\bibitem[\protect\citeauthoryear{Aller et~~al.}{1996}]{Allerbook1996}
  Aller L.~H., Appenzeller I., Baschek B., et~al., 1996, 
  Landolt-B\"ornstein: Numerical Data and Functional Relationships in
  Science and Technology —New Series “Gruppe/Group 6 Astronomy and
  Astrophysics” Volume 3 Voigt: Astronomy and Astrophysics. Extension
  and Supplement to Volume 2” Stars and Star Clusters (Berlin: Springer)

\bibitem[\protect\citeauthoryear{Bath}{1975}]{bath}
  Bath G. T., 1975, MNRAS, 171, 311

\bibitem[\protect\citeauthoryear{Bath \& Pringle}{1981}]{BathPringle81}
  Bath G. T. \& Pringle J. E., 1981, MNRAS, 194, 967

\bibitem[\protect\citeauthoryear{Cardelli et~al.}{1989}]{Cardelliet1989}
  Cardelli J.~A., Clayton G.~C., \& Mathis J.~S., 1989, ApJ, 345, 245

\bibitem[\protect\citeauthoryear{Fekel et~~al.}{2000}]{Fekelet2000}
  Fekel F.~C., Joyce R.~R., Hinkle K.~H., \& Skrutskie M.~F.,
  2000, AJ, 119, 1375

\bibitem[\protect\citeauthoryear{Fujimoto}{1982}]{Fujimoto1982}
  Fujimoto M.~Y., 1982, ApJ, 257, 767

\bibitem[\protect\citeauthoryear{Hameury et~al.}{2020}]{Hameuryet2020}
  Hameury J.-M., Knigge C., Lasota J.-P., Hambsch F.~J., \& James R.,
  2020, A\&A, 636, A1

\bibitem[\protect\citeauthoryear{Ilkiewicz et~al.}{2023}]{Ilkiewiczet2023}
Iłkiewicz K., Mikołajewska J., \& Stoyanov K.~A., 2023, ApJL, 953, L7

\bibitem[\protect\citeauthoryear{Kippenhahn \& Weigert}{1991}]{Kippenhahn}
  Kippenhanh R., Weigert A., 1991, Stellar Structure and Evolution
  (Berlin: Springer-Verlag)

\bibitem[\protect\citeauthoryear{Lasota}{2001}]{Lasota2001}
  Lasota, J.-P. 2001, New Astron.\ Review, 45, 449

\bibitem[\protect\citeauthoryear{Luna et~al.}{2018}]{Lunaet2018}
  Luna G.~J.~M., Mukai K., Sokoloski J.~L., et~al., 2018, A\&A, 619, A61

\bibitem[\protect\citeauthoryear{Luna et~al.}{2020}]{Lunaet2020}
  Luna G.~J.~M., Sokoloski J.~L., Mukai K., Kuin M.~N.~P., 2020,
  ApJ, 902, L14

\bibitem[\protect\citeauthoryear{Munari et~al.}{2016}]{Munariet2016}
  Munari U., Dallaporta S., Cherini G., 2016, New Astron., 47, 7

\bibitem[\protect\citeauthoryear{Munari}{2023}]{Munari2023}
  Munari U., 2023, RNAA, 7, 145

\bibitem[\protect\citeauthoryear{Paczynski}{1983}]{Paczynski1983}
  Paczynski B., 1983, ApJ, 264, 282

\bibitem[\protect\citeauthoryear{Planquart et~al.}{2025}]{Planquartet2025}
  Planquart L., Jorissen A., \& Van Winckel H., 2025, A\&A, 694, A85

\bibitem[\protect\citeauthoryear{Prialnik}{1986}]{Prialnik1986}
  Prialnik D., 1986, ApJ, 310, 222

\bibitem[\protect\citeauthoryear{Schaefer}{2010}]{Schaefer2010}
  Schaefer B.~E., 2010, ApJS, 187, 275

\bibitem[\protect\citeauthoryear{Schaefer}{2019}]{Schaefer2019}
  Schaefer B.~E., 2019, AAS Meeting, 51, 122.07

\bibitem[\protect\citeauthoryear{Schaefer}{2023a}]{Schaefer2023a}
  Schaefer B.~E., 2023a, JHA, 54, 436

\bibitem[\protect\citeauthoryear{Schaefer}{2023b}]{Schaefer2023b}
  Schaefer B.~E., 2023b, MNRAS, 524, 3146

\bibitem[\protect\citeauthoryear{Schaefer et~al.}{2023}]{Schaeferetal2023}
  Schaefer, B.~E., Waagern E.~O., and the AAVSO Observers, 2023,
  The Astronomer's Telegram, 16107, 1

\bibitem[\protect\citeauthoryear{Shen \& Bildsten}{2009}]{ShenBildsten2009}
  Shen K.~J., Bildsten L., 2009, ApJ, 692, 334

\bibitem[\protect\citeauthoryear{Schlafly \& Finkbeiner}{2011}]
{SchlaflyFinkbeiner2011}
  Schlafly  E.~F., \& Finkbeiner D.~P., 2011, ApJ, 737, 103

\bibitem[\protect\citeauthoryear{Schlindwein \& Baptista}{2024}]{SB2024}
  Schlindwein W. \& Baptista R., 2024, ApJ, 975, 92

\bibitem[\protect\citeauthoryear{Schlindwein et~al.}{2025}]{sbl25}
  Schlindwein W., Baptista R., \& Luna G. J. M., 2025, ApJ, 989, 78

\bibitem[\protect\citeauthoryear{Selvelli et~al.}{1992}]{Selvelliet1992}
  Selvelli P.~L., Cassatella A., Gilmozzi R., 1992, ApJ, 393, 289

\bibitem[\protect\citeauthoryear{Shakura \& Sunyaev}{1973}]{ss73} 
  Shakura N. I. \& Sunyaev R. A., 1973, A\&A, 24, 337

\bibitem[\protect\citeauthoryear{Stanishev et~al.}{2004}]{Stanishevet2004}
  Stanishev V., Zamanov R., Tomov N., \& Marziani P., 2004, A\&A, 415, 609

\bibitem[\protect\citeauthoryear{Stoyanov et~al.}{2024}]{Stoyanovetal2024}
  Stoyanov K.~A., Luna G.~J.~M., Zamanov R., et~al., 2024, Bulgarian Astr.\
  J., in press (ArXiv: 2406.01971)

\bibitem[\protect\citeauthoryear{Warner}{2003}]{Warner2003}
  Warner B., 2003, Cataclysmic Variable Stars, Cambridge Astrophysics
  Series 28, (Cambridge: Cambridge University Press)

\bibitem[\protect\citeauthoryear{Wolf et~al.}{2013}]{Wolfetal2013}
  Wolf W.~M., Bildsten L., Brooks, J., Paxton B., 2013, ApJ, 777, 136

\bibitem[\protect\citeauthoryear{Woodward et~al.}{2023}]{Woodwardetal2023}
  Woodward C.~E., Banerjee D.~P.~K., Evans A., 2023, The Astronomer's
  Telegram, 16120, 1

\bibitem[\protect\citeauthoryear{Zamanov et~al.}{2023}]{Zamanovet2023}
  Zamanov R., Boeva S., Latev G.~Y., et~al., 2023, A\&A, 680, L18

\bibitem[\protect\citeauthoryear{Zamanov et~al.}{2024}]{Zamanovet2024}
  Zamanov R., Stoyanov K.~A., Marchev V., et~al., 2024, Astron.\,Nachr.,
  345, 20240036

\end{thebibliography}
\end{document}